# A unified transition mechanism from shock to detonation waves

Hao Yan[a], Haochen Xiong[a], Xin Han[a], Chongguang Shi[a], Yancheng You[a,*]

[a] School of Aerospace Engineering, Xiamen University, Xiamen 361102, PR China

Abstracts: The transition of shock-to-detonation is of great significance for the investigation of supernova formation, disaster prevention and supersonic propulsion technology. In this paper, the influence Equation of shock-to-detonation transition is summarized for the oblique detonation problem from aerodynamic analysis. The Equation integrates the effects of parameters such as chemical reaction, shock intensity and wall conditions, which quantitatively explains the physical mechanism of shock-to-detonation transition in the form of mathematical expression. Comparison with numerical simulation results as well as their gradients verified the reliability of the influence Equation. Further, the influence Equation can also be used to predict the critical conditions for the transition from shock to detonation transition form. In addition to oblique detonation, the influence Equation is compatible with the deflagration-to-detonation problem for normal detonation, which shows a wide applicability.

1. Introduction

Shock [1] and detonation [2] waves are two important types of physical phenomena in aerodynamics, which are widely distributed in supernova explosions [3], nuclear explosions, vehicle propulsion schemes [4] and many other aspects. Generally speaking, detonation is an extreme combustion phenomenon in which chemical reactions are induced by shock waves [5], which is characterized by high thermal cycle efficiency, supersonic flame propagation, and extremely short reaction time. The respective characteristics of the two and their interrelationships have been the focus of academic research, which is of great significance in figuring out the formation of cosmic planets, disaster prevention and hypersonic propulsion. In the two-dimensional plane, shock and detonation can be mainly divided into two categories according to their morphology: normal shock/ detonation and oblique shock/ detonation [6]. There are both interconnections and significant differences between the two types of shock/detonation, so the following two types detonation waves are introduced separately.

In terms of normal detonation, Chapman [7] and Jouguet [8] respectively obtained the stable propagation of normal detonation velocity according to the conservation law by assuming that the chemical equilibrium state is reached after the detonation wave, and the research is called CJ theory. The detonation velocity obtained by CJ theory is in good accordance with the experimental observation results, and it is one of the main basic theories of detonation research. However, the CJ theory ignores the specific chemical reaction process in the detonation and only regards the detonation wave as a shock with energy release. In order to further investigate the specific structure of the detonation wave, Zel'dovich [9], Von Neuman [10] and Doring [11] each independently proposed a structural model that the detonation wave is composed of induced excitations and a certain length of chemical reaction zone, which is called the ZND model. Deflagration to detonation is one of the main ways to form a normal detonation wave and has attracted a lot of attention. Liu [12] analyzed the aerodynamic relationship between combustion, shock and detonation wave in deflagration-to-detonation process, and derived the formula of the critical incoming Mach number that can be formed in the detonation. Comparison with the experimental results verified that the 0.64 CJ Mach number is the minimum Mach number for the formation of detonation for ethane air with an equivalence ratio of one. Similarly, Alexei et al. published a paper in Science to derive the critical conditions for deflagration-to-detonation based on ground-based experiments and numerical simulations, and applied them to supernova explosions of type Ia.

In terms of oblique detonations, it is known from the current studies that the transition from shock to detonation wave exists in both smooth and abrupt types. Visually speaking, for the smooth type, the angular difference between the shock and the detonation is small, and there is no obvious main transverse wave structure in the flow field. On the contrary, for the abrupt transition, the angular difference is large, and there is an obvious main transverse wave. Miao [13] points out that the pressure ratio between the post-wave of shock and detonation is the main reason for the formation of the main transverse wave, and based on which, he puts forward a smooth/abrupt transition criterion. By changing the wedge angle, incoming Mach number and equivalence ratio, different numerical simulation results are obtained, which all imply that the pressure ratio of 1.3 is a significant boundary between smooth and abrupt detonation structures. Wang [14] proposes the criterion of post-wave velocity ratio to the CJ velocity based on the definition of the overdriven degree. The numerical simulation results show that the velocity ratio of 1 can effectively differentiate between the two types detonation structures. Teng [6] more intuitively takes the angle difference between the shock and detonation as a criterion. and the results show that 15-18 degrees is the smooth/abrupt transition interval.

Based on the above studies, it should be recognized that the transition from shock to detonation is of great physical significance and has been a hot issue in academic research. However, the present studies lack a unified physical formula to describe the transition process of normal/oblique shock and detonation. The reason is due to the fact that the combustion process of detonation is very complex, and different fuels with different incoming flow conditions and different geometrical constraints will have an impact on the detonation wave flow field, bringing about non-constant and non-uniform characteristics. In this paper, inspired by the CJ theory, the complex chemical reaction process of detonation combustion is attributed to the energy release, and the transition process of shock to detonation is investigated on this basis. The article firstly describes the physicochemical process and the governing Equations of the transition from shock to detonation, and then introduces different types of discriminants. Based on the summarization and analysis of previous studies, this article proposes the influence factor Equations that can quantitatively describe the shock/ detonation transition. Comparison with numerical simulation results verifies the rationality of the equation.

2. Physical models and governing equations

Considering that the oblique detonation has a more complex structure, this paper mainly focuses on it. The following is about wave system structures and aerodynamic parameters of oblique detonation.

2.1 Description of the problem

In the detonation flow field induced by oblique shock waves as depicted in Figure 1 (*a*), key wave structures include the oblique shock wave, the detonation wave, deflagration waves, and the transverse wave. Where the oblique shock wave indicated by solid blue line and the detonation wave represented by solid black line. Two transition types exist: smooth and abrupt type. Intuitively, when the difference in wave angles is small, the transition tends to be smooth; for instance, the shock wave remains unchanged while the detonation wave shifts downward to the white dashed line. Conversely, when the shock wave remains unchanged and the detonation wave shifts upward to the green dashed line, the transition is abrupt. More specifically, the presence of transverse waves indicated by solid yellow lines can generally serve as a criterion for distinguishing between the two transition modes. Smooth transitions lack transverse waves, whereas abrupt transitions involve them. Transverse waves are generally attributed to the mismatch in post-shock states between the upper and lower flow fields, resulting in pressure differentials that necessitate their involvement in balancing the flow field. In the lower region, the flow experiences temperature and pressure increase after passing through oblique shock waves, resulting in a series of deflagration waves downstream. In the upper region, the airflow directly goes through detonation waves.

The aerodynamic relationship of detonation waves is illustrated in Figure 1 (*b*), where incoming conditions and wedge angles directly influence the angle of detonation waves. The angle of detonation waves, in turn, interacts with incoming conditions to affect energy release, which further influences the wave angle, demonstrating a coupled effect. These four parameters collectively influence the aerodynamic parameters of detonation waves, which directly impact the type of shock/detonation wave transition. In fact, whether based on the structure of detonation flow fields or aerodynamic parameter relationships, the essence of the two transitions lies in the "gap" between shock waves and detonation waves. To quantitatively measure this difference and provide criteria for the two transitions, various researchers have proposed different criterions.

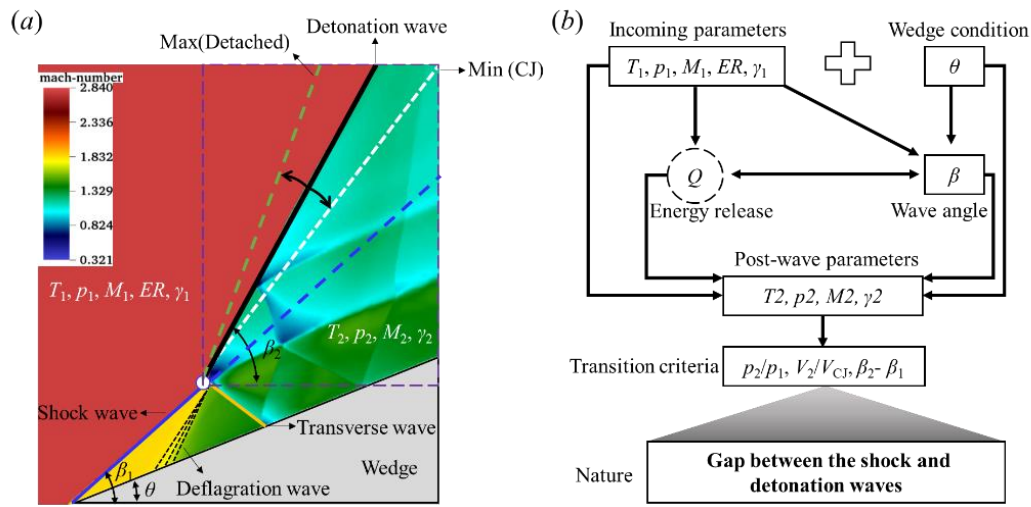

Figure 1. Schematic diagram of oblique shock/detonation waves: (*a*) wave system structure of the flow field based on numerical simulation [13], blue solid line is shock wave and blue dashed line is its extension line, black line is the detonation wave and two different white dashed line represents the stationary window under the same wedge angel, yellow line is the transverse wave. (*b*) relationship between aerodynamic parameters and transition criteria.

2.2 Introduction to the various types of criteria

Miao [13] pointed out that the ratio of post-wave pressures of oblique shock and detonation waves can serve as an indicator for measuring the abrupt degree of oblique detonation. According to his numerical simulation results, the ratio of 1.3 between these two pressures is a clear threshold, as shown in Equation (1).

$$\frac{p_d}{p_s} = 1.3 \quad (1)$$

Where, $p_d$ represents the pressure after the oblique shock wave, and $p_s$ denotes the pressure after the oblique shock. When the pressure ratio exceeds 1.3, the flow field exhibits an abrupt transition; conversely, when the pressure ratio is less than 1.3, the flow field shows a smooth transition. The physical explanation of this criterion is clear. When the pressure after the detonation significantly exceeds that after the shock wave, a primary transverse wave appears between them to balance the pressure difference between the two regions. This transverse wave further leads to the appearance of a triple wave point (shock wave, detonation wave, and primary transverse wave) in the flow field, which is typically regarded as a hallmark of the abrupt type. This criterion has a very clear physical explanation, but the threshold of 1.3 relies too heavily on specific numerical simulation results and cannot be universally generalized.

Furthermore, Han [15] noted in earlier research that the presence or absence of a subsonic region behind the wave can serve as a criterion for judging the smooth or abrupt structure of detonation waves, as shown in Equation (2).

$$M_2 = 1. \quad (2)$$

Specifically, in his numerical simulation results, the smooth type detonation waves are all followed by supersonic speeds, and the subsonic region exists behind the abrupt type detonation waves. Considering that subsonic velocities produce disturbances to the upstream, this criterion and its physical interpretation can provide support for the phenomenon that abrupt structures are more unstable. In an earlier study, Wang [14] used the ratio of the post-wave velocity to the CJ velocity as a discriminant for determining whether a transverse wave is present or not, as shown in Equation (3).

$$\Phi = \frac{V_2}{V_{CJ}} = 1. \quad (3)$$

Here, $V_2$ represents the post-wave velocity, while $V_{CJ}$ denotes the Chapman-Jouguet velocity of the detonation wave. According to his numerical simulations, when this ratio is less than one, transverse waves exist; otherwise, they do not. The physical significance of this criterion is that it can be viewed as an overdriven degree of the oblique detonation, which is also consistent with the laws inherent in Han's study [15].

Considering that accurate post-wave parameters usually require detailed numerical simulations to be obtained, it is not possible to make a timely judgment of the flow field for a given incoming flow condition. In order to overcome this weakness, Miao [13] gives another criterion for judging the structure of the transition zone based on the ratio of the incoming velocity and the CJ velocity by observing the pattern of the numerical simulation results, as shown in Equation (4):

$$\phi = \frac{V_1}{V_{CJ}} = 1.38. \quad (4)$$

Where $V_1$ is the velocity of the incoming flow. The physical mechanism of this criterion lies in the fact that, for oblique detonations, the larger the CJ velocity, the greater the ability of the detonation wave to propagate forward, i.e., the greater the angle of the detonation wave, for a given velocity of the incoming flow. Conversely, the smaller the CJ velocity, the greater the ability of the detonation wave to propagate downstream, the smaller the angle of the detonation wave. Therefore, this velocity ratio can be used as a performance of detonation wave angle size of a judgment and thus determine the structure of the transition zone of the detonation. In most of the examples he showed, 1.38 is a clear threshold. A similar criterion was also presented in Teng's study [6], with the difference that Teng used the Mach number instead of the velocity ratio:

$$M_{cr} = \frac{M_{in}}{M_{CJ}} = 1.44. \quad (5)$$

In addition, Teng further pointed out that the difference between the angle of the detonation wave and the angle of the shock wave can be used as an intuitive criterion, as shown in Equation (6). This criterion is very intuitive. At a small difference between the angle of the detonation wave and the shock, the transition zone shows a smooth structure, and vice versa shows an abrupt transition structure.

$$\beta_2 - \beta_1 = 15° - 18°. \quad (6)$$

where $\beta_2$ is the detonation wave angle and $\beta_1$ is the shock angle. In different examples, 15° to 18° is the main threshold range for the smooth/abrupt transition. The applicability of this criterion is demonstrated through different examples. The physical mechanism involved is that when the angular difference between the two is small, the compression wave is sufficient to complete the angular transition and will therefore be a smooth structure. Otherwise, the abrupt structure is necessary to fulfill the larger angular difference.

3. Transition mechanisms of oblique detonation waves

All the above many criterions have their own existing physical mechanisms, and all of them have been verified by their own numerical simulation results, so they can be considered reasonable in their own studies. However, so many criterions also make the problem confusing. Which parameters are the key factors determining the structure of the transition zone of detonation waves, and does a clear theoretical explanation exist? Are there hidden connections between the various criterions? In addition, logically, although the transition zone structure of detonation waves is very much related to the post-wave state (such as transverse wave and subsonic region, etc.), the real parameters which determinants the smooth/abrupt type should be the incoming flow parameters and wall conditions. In order to resolve the above confusion, the above criterion was further developed and analyzed.

3.1 Aerodynamic parameters in the transition from shock to detonation waves

Analyzing the multiple criterions mentioned above, it is easy to find that the study mainly focuses on the aerodynamic parameters such as post-wave pressure, velocity, and wave angle. Therefore, this paper analyzes the involved aerodynamic parameters one by one. First of all, the post-wave pressure of shock and detonation can be solved through the RH relationships. For the oblique shock wave:

$$\frac{p_s}{p_1} = 1 + \frac{2\gamma_1}{\gamma_1+1}(M_1^2 \sin^2\beta - 1). \quad (7)$$

Similarly, the post-wave pressure of the detonation can be obtained as:

$$\frac{p_d}{p_1} = 1 + \gamma_1 M_1^2 \sin^2\beta(1-X),$$

$$X = \frac{1 + \gamma_1 M_1^2 \sin^2\beta \pm \sqrt{(M_1^2 \sin^2\beta - 1)^2 - 2(\gamma_1+1)M_1^2 \sin^2\beta \tilde{Q}}}{(\gamma_1+1)M_1^2 \sin^2\beta}. \quad (8)$$

Where $\tilde{Q}$ is the dimensionless energy release, calculated as shown in Equation (9):

$$\tilde{Q} = Q/c_p T_1. \quad (9)$$

Where $c_p$ is the constant pressure-to-heat ratio and $T_1$ is the incoming flow temperature. Thus, the ratio of the post-wave pressures of the detonation and the shock is:

$$\frac{p_d}{p_s} = \frac{1 + \gamma_1 M_1^2 \sin^2\beta(1-X)}{1 + 2\gamma_1/(\gamma_1+1)(M_1^2 \sin^2\beta - 1)}. \quad (10)$$

It is clear that the ratio of pressure of the detonation to shock is related to the four variables of the incoming Mach number, specific heat ratio, wave angle, and energy release, i.e:

$$\frac{p_d}{p_s} = f(\gamma_1, M_1, \beta, \tilde{Q}). \quad (11)$$

In a similar analytical manner, the post-wave velocity of the detonation wave is analyzed:

$$V_2 = \frac{V_1 \cos\beta}{\cos(\beta-\theta)}. \quad (12)$$

For the CJ velocity, Lee [5] gives an approximate solution:

$$M_{CJ} \approx \sqrt{2(\gamma_1^2-1)\bar{Q}}. \quad (13)$$

The CJ velocity is related to the specific heat ratio and energy release, and the CJ Mach number is also related to the local speed of sound. Thus, the ratio of post-wave velocity to CJ velocity can be expressed as:

$$\Phi = \frac{V_2}{V_{CJ}} = \frac{V_1 \cos\beta}{\sqrt{2(\gamma_1^2-1)\bar{Q}} \cos(\beta-\theta)}. \quad (14)$$

Similarly, it can be known that the ratio of velocities is also related to the four variables of incoming velocity, specific heat ratio, wave angle and energy release, i.e.:

$$\Phi = f(\gamma_1, V_1, \beta, \tilde{Q}). \quad (15)$$

Further, it is known from Teng's study [6] that the relationship between the shock and detonation wave angle needs to be expressed as an implicit function:

$$\frac{\tan \beta_1}{\tan(\beta_1 - \theta)} = \frac{(\gamma+1)M_1^2 \sin^2 \beta_1}{2+(\gamma-1)M_1^2 \sin^2 \beta_1}. \quad (16)$$

Where the angle of the detonation wave can be obtained by determining the relation Equation:

$$\theta = \beta_2 - \arctan \frac{\tan \beta_2}{X} \quad (17)$$

Therefore, the difference between the angle of the shock and the detonation wave can be known:

$$\beta_2 - \beta_1 = f(\gamma_1, M_1, \theta, \tilde{Q}) \quad (18)$$

Overall, it can be known based on the above analysis:

$$\frac{p_d}{p_s} = f(\gamma_1, M_1, \theta, \tilde{Q}), \Phi = \frac{V_2}{V_{CJ}} = f(\gamma_1, V_1, \bar{Q}, \theta), \beta_2 - \beta_1 = f(\gamma_1, M_1, \theta, \tilde{Q}). \quad (19)$$

Based on the above summary, the individual criterions are ultimately shown to be related to four main variables, the incoming Mach number (velocity), the incoming specific heat ratio, the wedge angle and the energy release. Among them, the first three are generally more fixed and easier to determine, but the energy release is related to the specific fuel equivalent ratio, detonation wave angle, the temperature of the incoming flow, pressure and many other parameters, it is impossible to be expressed as a function of a definite relationship. Generally speaking, the energy release can be obtained by numerical simulation or chemical equilibrium method, and its influencing factors are shown in Equation (20):

$$\tilde{Q} = f(ER, M_1, T_1, p_1, \beta, ...). \quad (20)$$

According to the above analysis of aerodynamic parameters, it can be known that all types of criterions are mainly related to four types of parameters, so this paper gives the following mechanism factor equation to measure the gap between shock and detonation waves:

$$\eta = \frac{Q^2}{\gamma_1 M_1^2 \sin \theta}. \quad (21)$$

In this mechanism factor equation, the reason why energy release is in the form of a square rather than linear is that it is considered that energy release not only affects the parameters of the post-wave directly, but also affects the post-wave parameters indirectly by affecting the wave angle. Therefore, the squared form is given here. To clarify, this $\eta$ should be understood as the key influence factor of the transition mechanism rather than the criterion.

3.2 Verification of the mechanism factor equation

In order to verify the rationality of the above mechanism factor equation, this paper compares the mechanism factor $\eta$ with several criterions mentioned above, and the results are shown in Figure 2.

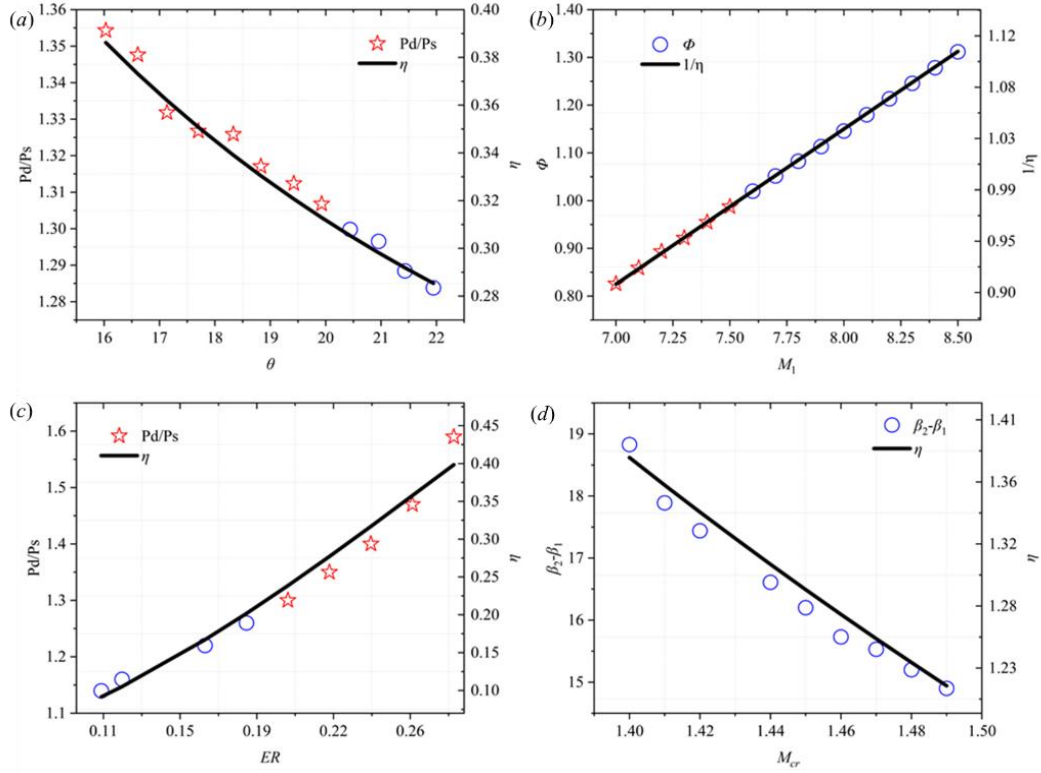

Figure 2. The comparison results of the influence factors with each criterion: (*a*) the effect of wedge angle variation on smooth/abrupt type is investigated with the pressure ratio criterion, (*b*) the effect of incoming Mach number variation on smooth/abrupt type is investigated with the overdrive criterion, (*c*) the effect of equivalence ratio variation on smooth/abrupt type is investigated with the pressure ratio criterion, (*d*) the effect of incoming Mach number variation on smooth/abrupt type is investigated with the wave angle difference as a criterion.

According to Figure 2, the mechanism factor proposed in this paper shows a strong similarity with the individual criterions, which implies the reasonableness of the mechanism factor to some extent. It is to be noted separately that in Figure 2 (*b*), since $\Phi$ unlike the other criterions, the larger its value is, the smoother the transition of the detonation wave is. Therefore, it will be compared with $1/\eta$. In addition, the rationality of the mechanism factor equation can also be further verified by observing the gradient variation rule of the numerical simulation results with the independent variables. Specifically, in Figure 2 (*a*), for example, the magnitude of the change in the pressure ratio gradually slows down as the wedge angle increases. And in Figure 2 (*c*), the change amplitude of the pressure ratio gradually increases with the increase of the equivalence ratio. Meanwhile, the derivation of the influence Equation (21) leads to Equation (22):

$$\frac{d\eta}{dQ} = \frac{2Q}{\gamma_1 M_1^2 \sin\theta}, \frac{d\eta}{dM_1} = \frac{-2Q^2}{\gamma_1 M_1^3 \sin\theta}, \frac{d\eta}{d\theta} = -\frac{Q^2}{\gamma_1 M_1^2 \sin\theta \tan\theta}. \quad (22)$$

According to the above Equation (22), it can be seen that the derivative increases gradually as the energy release (equivalence ratio) increases, which implies that this mechanism factor is gradually increasing. Moreover, as the Mach number increases, the value of its conductance then decreases gradually. Similarly, in the range of 0 to 90 degrees, the value of its derivative decreases gradually as the wedge angle increases. The gradient changes revealed by the above equations are in agreement with the numerical simulation results. This again justifies the mechanism factor proposed in this paper.

In general, a sound physics formula should not only be able to explain existing laws but also need to make predictions about unknown phenomena. For example, the law of gravity not only explains why an apple falls to the ground, but also successfully predicts the existence of Neptune. In addition to the validation of the laws in comparison with the numerical simulation results, the mechanism factor equation proposed in this paper should also be able to predict the conditions of smooth/abrupt transitions.

As shown in Figure 3 (*a*), it is known from the mechanism factor equation that an abrupt detonation can be transformed to a smooth one by increasing the incoming Mach number or decreasing the equivalence ratio. Since the mechanism factor equation can quantify the change in the degree of influence due to the change in conditions, these two changes can theoretically be converted to each other. Specifically, a change in the equivalence ratio that has the same effect as a change in Mach number can be obtained through the calculation of the mechanism factor equation. To realize this, as shown in Figure 3 (*b*), firstly, an abrupt oblique detonation result is chosen with an incoming velocity of $V_1$ and an equivalence ratio of $ER_1$ under a certain wedge angle. Next, the critical velocity $V_2$ that transforms into a smooth structure as the incoming velocity increases can be computed. The mechanism factor $\eta$ at this point can be computed according to the mechanism factor equation. The equivalence ratio corresponding to $\eta$ can be inverted and solved for the condition of $V_1$, $ER_2$. If the mechanism factor is correct, the structure corresponding to $ER_2$ should also be a critical smooth type oblique detonation wave. Therefore, the reasonableness of the mechanism factor equation can be proved by observing whether the numerical simulation results under the conditions of $V_1$ and $ER_2$ are consistent with the predicted results.

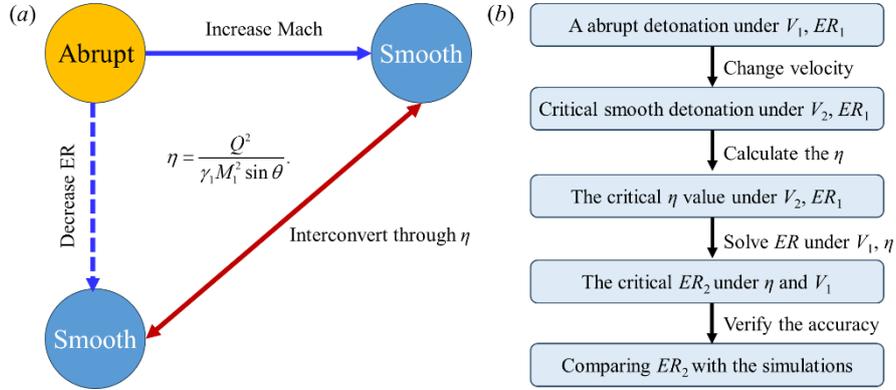

Figure 3. Critical conditions for smooth/abrupt transitions are predicted by means of the criterions proposed in this paper: (*a*) schematic diagram of the validation method, (*b*) flowchart of the validation. In the case of the Miao's [13] study, for example, an initial standard state is first selected:

$$V_1 = 1690 \text{m/s}, ER=0.25, \theta=22°. \quad (23)$$

Currently, the oblique detonation is an abrupt type, and when the incoming velocity increases to 1900 m/s, the oblique detonation becomes a critical smooth type. According to the proposed in this paper can be calculated at this time $\eta$ is 0.2045, if the initial standard state does not change the incoming velocity but to change the equivalence ratio, the smooth type can also be achieved. According to the mechanism factor equation proposed in this paper, the critical equivalence ratio of the smooth type structure can be solved to be 0.2 for a velocity of 1690 under $\eta$ (0.2045), which is in good agreement with the numerical simulation results. This also proves again the rationality of the mechanism factor proposed in this paper.

Table 1. Comparison of the theoretical results calculated based on the mechanism factor equations with the numerical simulation results

|  | Energy release $Q_2$ | Equivalence ratio $ER_2$ |
|---|---|---|
| Theoretical results by $\eta$ | 0.89 | 0.20 |
| Numerical simulation results | 0.84 | 0.19 |
| Error | 6% | 5% |

3.3 Application in DDT

In the case of a normal detonation, Equation 21 degenerates into:

$$\eta = \frac{Q_{CJ}}{\gamma_1 M_1^2}. \quad (24)$$

The DDT process is standardized on the CJ state, so the energy release of the fuel and the post-wave Mach number can be determined. According to the mechanism factor equation, it is known that the numerator is determined and the smaller the denominator, the larger the whole. Therefore, there must be a lower limit for the Mach number, otherwise the gap between the shock and detonation will be too large. This is also consistent with the conclusion of Liu's study [12]:

$$\frac{6M_1^2}{M_1^2+5} = \frac{7M_{CJ}^2}{6}\frac{T_1}{T_0}. \quad (25)$$

Where the incoming static temperature and the total combustion temperature represent the energy release. In addition to the above analysis and application, the mechanism factor equation proposed in this paper can also explain the effect of the incoming temperature change on the transformation. Under the same conditions, the higher the temperature of the incoming flow, the less energy can be released, and the smaller the gap between the shock and the detonation will be, and the transition will be to a smooth type.

4. Conclusion

1) By analyzing the flow field structure and governing equations of the detonation, this paper summarizes the transition mechanism factor of shock to detonation waves and puts forward the equation of the mechanism factor.
2) Combined with multiple criteria and their numerical simulation results, the rationality of the mechanism factor equation is comprehensively verified from the perspective of the zero-order parameter and the first-order gradient. Further, the equation can be used to predict the critical conditions of the smooth/abrupt transition and agrees well with the numerical simulation results.
3) The mechanism factor equation of oblique detonation can be compatible with the DDT process for normal detonations which can provide theoretical analysis tools for DDT detonation conditions.

Declaration of Interests

The authors report no conflict of interest.

Acknowledgements

1) The authors acknowledge the support of the National Natural Science Foundation of China (Grant Nos. U20A2069, U21B6003, and 12302389) and the Advanced Aero-Power Innovation Workstation (Grant No. HKCX2024-01-017).
2) Discussions with Prof. Liu deepened the understanding of the DDT process in this paper, for which the authors are grateful.
3) Dr. Zijian Zhang's research proposed a chemical equilibrium method to accurately calculate the post-wave parameters and the energy release, which is facilitated by this paper.